\documentclass{aastex}          
\usepackage{spr-astr-addons}    

\begin{document}
%
  \title{Is radio jet power linearly proportional to the product of central black
hole mass and Eddington ratio in AGN?}

\shorttitle{Radio jet power, black hole mass and Eddington ratio}

\shortauthors{Liu \& Han}

\author{Xiang Liu\altaffilmark{1,2}}
\and
\author{Zhenhua Han\altaffilmark{1,3}}

\email{liux@xao.ac.cn}

\altaffiltext{1}{Xinjiang Astronomical Observatory, Chinese
Academy of Sciences, 150 Science 1-Street, Urumqi 830011, PR
China}

\altaffiltext{2}{Key Laboratory of Radio Astronomy, Chinese
Academy of Sciences, Nanjing 210008, PR China}

\altaffiltext{3}{Graduate University of the Chinese Academy of
Sciences, Beijing 100049, PR China}

\begin{abstract}
A model for the relation between radio jet power and the product
of central black hole (BH) mass and Eddington ratio of AGN is
proposed, and the model is examined with data from the literature.
We find that radio jet power positively correlates but not
linearly with the product of BH mass ($m$ in solar mass) and
Eddington ratio ($\lambda$), and the power law indices ($\mu$) are
significantly less than unity for relatively low accretion
($\lambda<0.1$) AGN, $P_{j}\propto (\lambda m)^{\mu}$, in the
radio galaxies and the Seyfert galaxies. This leads to a negative
correlation between radio loudness and $\lambda m$ for the low
luminosity AGN, i.e. $R\propto (\lambda m)^{\rho}$ with
$\rho=(7/6)\mu-1<0$, which may be attributed to a contribution of
BH spin to total jet power assuming that the spin induced jet is
gradually suppressed as the accretion rate increases. Whereas, for
the high-z quasars which often show the slope $\mu\geq1$, a
positive correlation between the radio loudness and disc
luminosity is predicted. We discuss that the jet powers of the
high-z FRII quasars are likely dominated by the accretion disc
rather than by the BH spin.
\end{abstract}


\section{Introduction}

It is generally accepted that active galactic nuclei (AGN) harbor
massive black holes (BHs). The black holes have only three
physical parameters: mass, spin and net charges, the net charges
are often considered to be zero. The mass (and its time
derivative) and spin (and its time derivative) of BHs are crucial
for understanding the AGN phenomena; most models and simulations
suggest that it is the mass accretion and/or spin of black holes
that produce the jets of AGN, e.g. see \citet{bla77} model for the
BH spin induced jet, \citet{bla82} model for the disc accretion
induced jet, and see recent reviews for the simulations of
accretion disc \citep{fra13} and for hot accretion flows
\citep{yuan14}. The co-evolution of central BH and its host AGN
makes it possible for measuring the BH mass via various ways, and
the mass accretion rate can be estimated through the measurements
of Eddington ratio. While BH spins are still difficult to measure,
although there is an explosion of BH spin measurements in recent
years, only some 19 AGN may have BH spin measurements
\citep{rey13}.

About 10\% of quasars are radio loud \citep{kal12}, and many more
are in radio weak/quiet state, while a higher radio loud fraction
was found in low luminosity AGN \citep{ho02}. It is not well
understood how a radio jet is related to the central BH mass,
spin, and accretion rate (or Eddington ratio) in AGN.
Observationally, radio jet powers are not well correlated with BH
masses \citep{ho02}, whereas there are several findings that radio
luminosity correlates with the narrow emission line luminosity or
the bolometric luminosity which assumed to be proportional to disc
accretion rate \citep{raw91, wil99, cao04, van13}. In recent
development of the \citet{bla77} mechanism, \citet{tch11} and
\citet{mck12} demonstrate that the magnetically arrested flows can
extract BH spin energy efficiently. A jet-spin correlation has
been suggested in X-ray binaries \citep{ste13}, but there are some
debates on the correlation \citep{rus13}. Furthermore, an
anti-correlation between the radio loudness and Eddington ratio is
found in relatively low luminosity AGN \citep{ho02, sik07, sik13},
which is probably due to the BH spin \citep{sik13, sikb13}. It is
likely that both the disc accretion and BH spin may contribute to
the jet power of AGN. In this paper, we propose a simple model
between radio jet power and disc accretion, and reanalyze the data
from the literature to examine the relation and discuss the
result.

\section[]{Relation of radio jet power to BH mass and Eddington ratio}

In the Newtonian approximation -- this is suitable to a distance
beyond a few Schwarzschild radii of BH \citep{mei12}, where a jet
may be formed from the disc accretion of AGN if not to consider a
contribution from the BH spin, the binding energy of unit mass in
a Keplerian orbit is $GM/2r$, where $G$ is the gravitational
constant. For an accreting mass of $\Delta{M}$ in time of
$\Delta{t}$, the binding energy $E$ per unit time is

\begin{equation}
E=\Delta{M}/\Delta{t}\times(GM/2r)=G \dot{M} M/2r.
\end{equation}

Where $\dot{M}$ is the accretion rate defined by accreted mass
$\Delta{M}$ per unit time. For the radiative efficiency of
$\varepsilon$ in an accretion disc, the disc luminosity:

\begin{equation}
L_{disc}=\varepsilon G \dot{M} M/2r.
\end{equation}

Similarly assuming radio jet power is also proportional to the
binding energy with a jet efficiency $\eta$, we will have the
radio jet power $P_{j}$:

\begin{equation}
P_{j}=\eta G \dot{M} M/2r.
\end{equation}

Considering the ratio of disc luminosity to the Eddington
luminosity:

\begin{equation}
\frac{L_{disc}}{L_{Edd}}=\frac{\varepsilon G \dot{M} M/2r}{4\pi G
M m_{p} c/\sigma_{T}}.
\end{equation}

We have

\begin{equation}
L_{disc}/L_{Edd}=5.28\times 10^{-12} \varepsilon \dot{M}/r.
\end{equation}

Taking this relation into equation (3), $r$ is cancelled, and we
have

\begin{equation}
P_{j}=1.26\times10^{38} (\eta /\varepsilon) [(L_{disc}/L_{Edd})
M/M_{\odot}] (erg/s).
\end{equation}

This relation shows that the jet power is the disc accretion
dominated, which is expected to be linearly proportional to the
product of the Eddington ratio ($\lambda=L_{disc}/L_{Edd}$) and
the BH mass ($M$) of AGN, with the coefficient of
$\eta$/$\varepsilon$. The jet power depends not only on the
Eddington ratio but also BH mass in this relation. It is
equivalent to the relation (from equations 2 and 3):

\begin{equation}
P_{j}=(\eta /\varepsilon)L_{disc}.
\end{equation}

\section[]{Statistical analysis of the jet power, BH mass and Eddington ratio}

To test the formula (6), we searched for suitable AGN samples
which have information on radio jet luminosity, BH mass and
Eddington ratio (or disc luminosity). Firstly, we study from the
\citet{sik13} sample, the sample consists of 404 narrow line radio
galaxies (NLRGs) consisting of FRIs and FRIIs, double-double radio
lobes, X-shaped lobes, and one side lobes (for details, see
\citealt{sik13}), and the sample is limited to redshift $<$0.4. We
use only the data of FRIs and FRIIs, they are the majority of the
sample. The 1.4~GHz radio luminosity computed from 1.4 GHz fluxes
in the NVSS catalog, BH mass and $H_{\alpha}$ (also $[O_{III}]$)
line luminosity are available in the Sikora sample, where the
black hole masses were estimated from the observed stellar
velocity dispersion given in the SDSS using the relation by
\citet{tre02} with typical error of log$M$ less than 0.3 dex. The
radio jet power $P_{j}$ can be estimated from low frequency (151
MHz) radio luminosity with the minimal energy argument for
synchrotron emission \citep{wil99}, assuming that the jet output
results in energy stored in radio source lobes together with
associated work done on the source environment. We convert the jet
power formula of \citep{wil99} from 151 MHz to 1.4 GHz assuming
source spectral indices $S_{\nu}\propto \nu^{-0.8}$, and we have

\begin{equation}
P_{j}=2.32\times10^{20} (f/3)^{3/2} (P_{1.4})^{6/7}[W/Hz] (erg/s)
\end{equation}

The $f$ (in range 1-20) represents several uncertainties
associated with estimating $P_{j}$ from 151~MHz luminosity
\citep{wil99}. We here use the median value $f=10$ \citep{blu00}
and use the equation (8) for that we want to study the power law
slope between $P_{j}$ and disc luminosity through $P_{1.4}$.
\citet{sik13} used a simplified formula $P_{j} \propto P_{1.4}$,
that overestimates a jet power by more than a factor of 10 than
that of equation (8). Where the $P_{1.4}$ is the 1.4 GHz
luminosity, the radio jet power $P_{j}$ is generally less than the
bolometric disc luminosity in our case. The bolometric luminosity
$L_{bol}$ estimated with the $H_{\alpha}$ line luminosity, the
same as in \citet{sik13}, see also \citet{net09}:

\begin{equation}
L_{disc}=L_{bol} \simeq 2000 L_{line} (erg/s)
\end{equation}

The Eddington ratio is computed from $L_{disc}/L_{Edd}$, and the
Eddington luminosity $L_{Edd}$ depends only on BH mass.

The radio loudness parameter $R$ is defined by the 1.4 GHz radio
luminosity over the $H_{\alpha}$ line luminosity, i.e.
$R=P_{1.4}/L_{H_{\alpha}}$.

To test and fit the equations (6)-(7), we rewrite the equation
(6)-(7) as a power law form $P_{j}=(\eta^{'}
/\varepsilon^{'})L_{disc}^{\mu}$, i.e.:

\begin{equation}
log(P_{j})=b + \mu \times log(\lambda M/M_{\odot}),
\end{equation}

where $b=log [1.26\times10^{38} (\eta^{'} /\varepsilon^{'})]$, the
ratio $\eta^{'}/\varepsilon^{'}$ is a coefficient for the power
law form and it returns to $\eta/\varepsilon$ when $\mu=1$.

For the anti-correlations between the radio loudness and disc
luminosity (or Eddington ratio) found by \citet{ho02},
\citet{sik07, sik13}, it is reasonable to assume a power law
relation, i.e: $R=\xi L_{disc}^{\rho}$, and rewrite it with the
form ($m\equiv M/M_{\odot}$):

\begin{equation}
log(R)=c + \rho \times log(\lambda m)
\end{equation}

\begin{figure}
    \includegraphics[width=8cm]{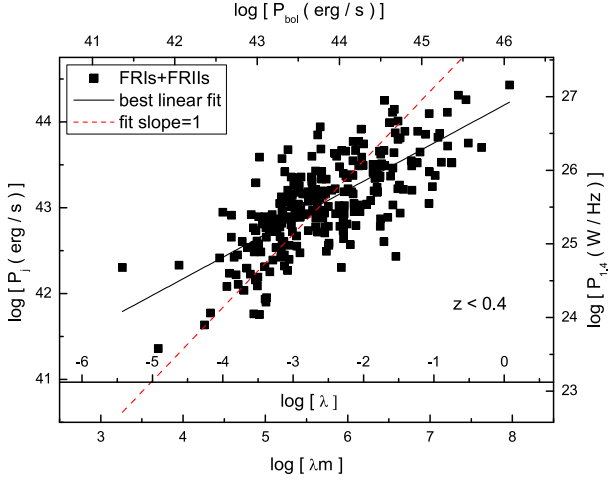}
    \caption{Log[jet power] vs. log[$\lambda m$]
    for the narrow line radio galaxies (FRIs+FRIIs), with the best linear fit to the sample, the dash line is fitted with the fixed slope=1.
    The top axis is the disc luminosity, the right axis is the 1.4 GHz luminosity, and the Eddington ratio is approximately marked in the inside X-axis. }
     \label{fig1}
  \end{figure}

\begin{figure}
    \includegraphics[width=8cm]{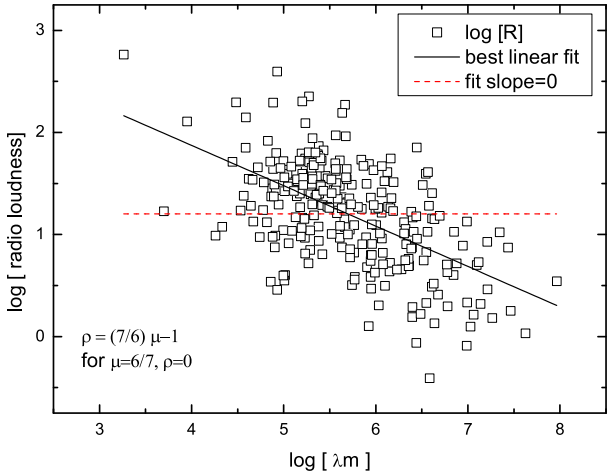}
    \caption{Log[radio loudness] vs. log[$\lambda m$]
    for the narrow line radio galaxies (FRIs+FRIIs), with the best linear fit to the sample, the dash line is fitted with the fixed slope=0.}
     \label{fig2}
  \end{figure}

Then, we plot the radio power $P_{j}$ versus the production of BH
mass and Eddington ratio for the sample, as well as the radio
loudness $R$ versus $\lambda m$, and fit the data with the
equation (10),(11) respectively, as shown in Fig.~\ref{fig1} and
Fig.~\ref{fig2}. The linear regression fitted parameters are
listed in Table~\ref{tab1}, as well as the correlation coefficient
and the null hypothesis probability in Table~\ref{tab2}.

\begin{figure}
    \includegraphics[width=8cm]{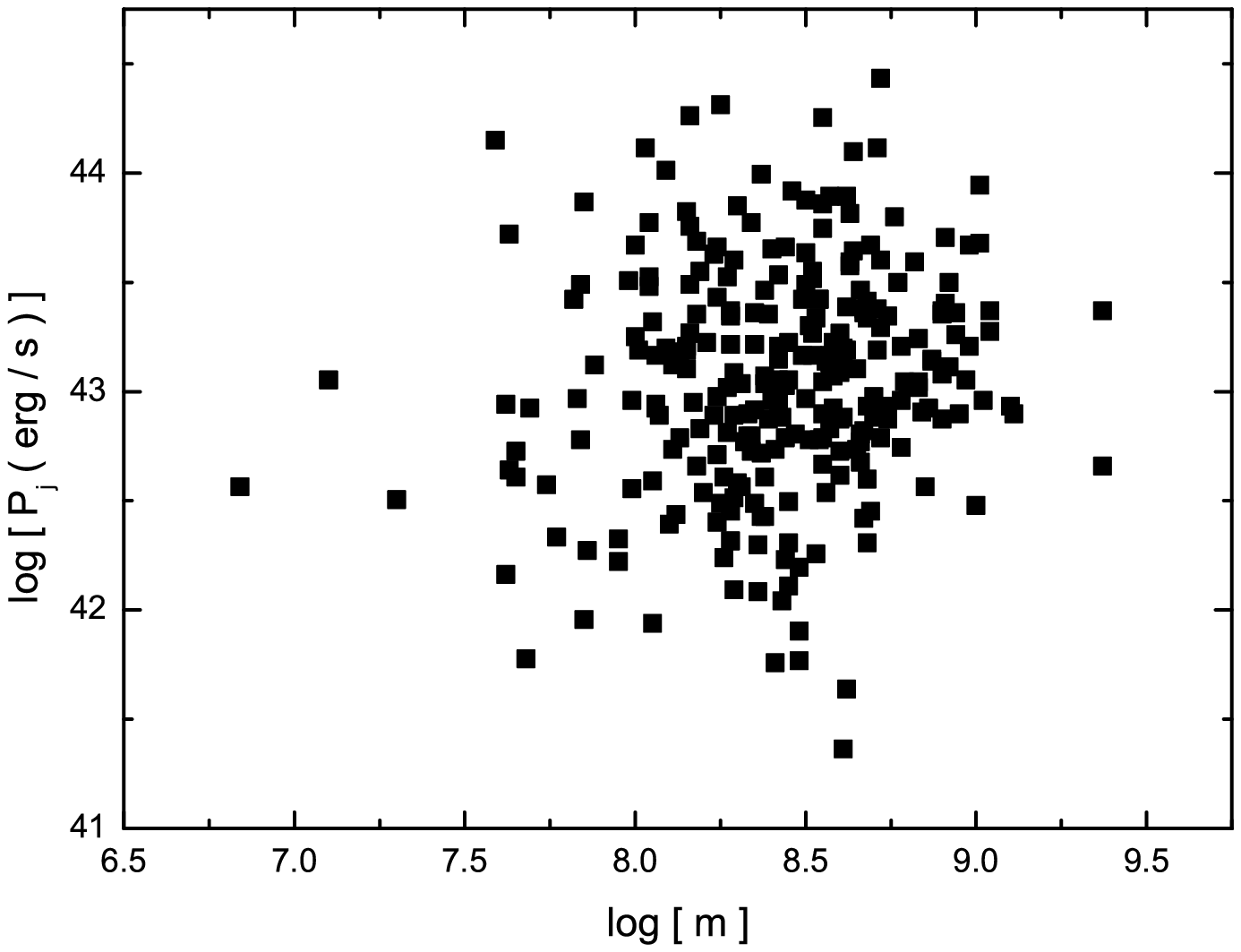}
    \caption{Log[jet power] vs. log[$m$] for the narrow line radio galaxies (FRIs+FRIIs), $m$ is black hole mass in solar mass unit.}
     \label{fig3}
  \end{figure}

The result shows that the radio power of the NLRGs (FRIs+FRIIs)
positively correlates with the $\lambda m$, with the power law
index of 0.52$\pm$0.06 with high correlation coefficient and
confidence level in Table~\ref{tab2}, it is not a linear
proportionality as one might expect from equations (6)-(7). The
radio loudness is anti-correlated with the $\lambda m$ with the
power law index of -0.40$\pm$0.08. It shows that the jet powers
have no correlation with the BH masses (Fig.~\ref{fig3}). If we
fit separately for FRIIs and FRIs in the sample, the slope $\mu$
is 0.52$\pm$0.08 and 0.49$\pm$0.18 for the FRIIs and FRIs
respectively.

We also study the \citet{sik07} sample, that consists of five
sub-samples including broad-line radio galaxies (BLRGs),
radio-loud quasars (RLQs), FRI radio galaxies, optically selected
quasars (PG quasars), and Seyfert galaxies plus LINERS (only 3),
all the sources are at redshift z$<$0.5. The 5~GHz radio
luminosity, BH mass and Eddington ratio are available in the
\citet{sik07} sample. We searched total 1.4 GHz flux densities
available from the literature (\citealt{con98}; and data from the
NED and references therein) for the sample, for that the lower
frequency radio luminosity would reflect the isotropic properties
of emission in the lobes and extended jets. The BH masses in
\citet{sik07}, were estimated using a broad-line region
size-luminosity relation, assuming virial velocities of the gas,
which produces broad $H_{\alpha}$ lines \citep{gre05}, or from the
appropriate references. The uncertainties of BH mass from
\citet{gre05}, typically $\sim$20\% in the virial mass formula
that depends on the $H_{\alpha}$ line alone. The Eddington ratio
is $\lambda$=$L_{bol}/L_{Edd}$, and the bolometric luminosity
$L_{bol}$ is assumed to be 10 times optical $B$-band nuclear
luminosity (at 4400$\AA$), i.e. $L_{bol}=10L_{B}$ (see, e.g.,
\citealt{ric06}), where $L_{B}=\nu_{B} L_{\nu_{B}}$, which is from
directly measured apparent magnitudes of the nuclear regions or
also estimated from the $H_{\alpha}$ line, the errors of
$H_{\alpha}$ flux can be $\leq$30\% \citep{gre05}, see
\citet{sik07} for more details on the estimation of BH mass and
nuclear disc luminosity. Here we use the equation (8) to estimate
the jet power by using $f=10$ and $P_{1.4}$.

\begin{figure}
    \includegraphics[width=8cm]{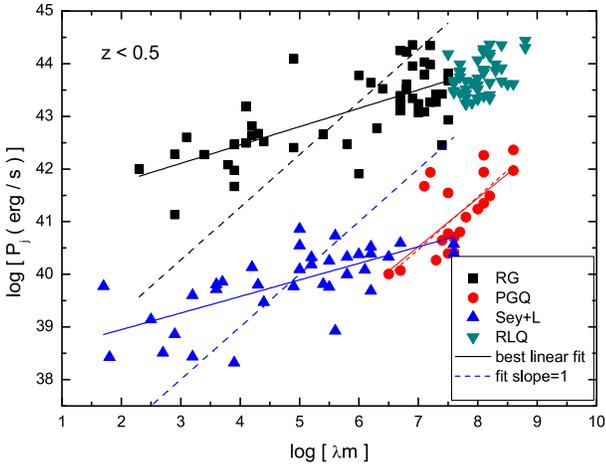}
    \caption{Log[jet power] vs. log[$\lambda m$]
    for the radio galaxies (RG), Seyfert
galaxies and LINERS (Sey+L), and PG quasars (PGQ), with the best
linear fit to each subsample, the dash line is fitted with the
fixed slope=1. }
     \label{fig4}
  \end{figure}

\begin{figure}
    \includegraphics[width=8cm]{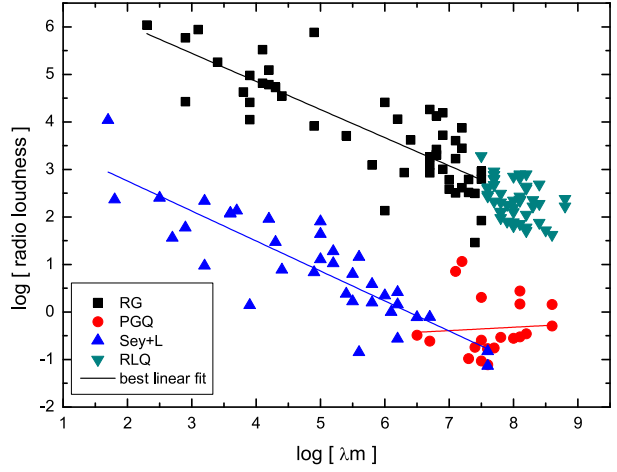}
    \caption{Log[radio loudness] vs. log[$\lambda m$]
    for the sub-samples (same as in Fig.4), with the best linear fit to each subsample.}
     \label{fig5}
  \end{figure}

As each subsample of the \citet{sik07} sample is relatively small,
to compare with the sample in Fig.~\ref{fig1} which are narrow
line FRI/FRII galaxies, we combine the FRI galaxies and BLRGs into
radio galaxies (RG) in the \citet{sik07} sample, also because the
FRI and FRII galaxies may have similar accretion mode
\citep{cao04}. The BLRGs and RLQs which divided by absolute
magnitude $M_{V}>-23$ and $M_{V}<-23$, are almost all the FRIIs
\citep{sik07}. We plot the jet power $P_{j}$ derived from 1.4 GHz
luminosity versus $\lambda m$ in Fig.~\ref{fig4} as well as the
radio loudness versus $\lambda m$ in Fig.~\ref{fig5}. A few
outliers are excluded in our analysis (e.g., the NGC1275 from
Seyfert galaxies, which is actually hosted by a giant elliptical
galaxy, and 5 PGQs which are in the RLQ area), and those data with
only upper/lower limits are excluded. The radio loudness is
computed with $R=1\times10^{5}L_{1.4}/L_{B}$ which converted from
the formula $R=1.36\times10^{5}L_{5.0}/L_{B}$ at 5 GHz in
\citet{sik07}, assuming source spectral index $S_{\nu}\propto
\nu^{-0.8}$. Linear regression fittings to the RGs,
Seyferts+LINERS, and PG quasars are shown in Fig.~\ref{fig4} with
the fitted parameters in Table~\ref{tab1}, as well as the
correlation coefficient and null probability in Table~\ref{tab2}.

The quality of a linear regression can be measured by the
coefficient of determination (COD), a value from 0 to 1. If it is
close to 0 the relationship between X and Y will be regarded as
very poor, the COD theoretically equals to the square of the
Pearson coefficient of linear correlation. We try to do linear
fitting with slope=1 (dash line in Fig.~\ref{fig1}), the resulted
COD is 0.07, which is much lower than the COD=0.52 of the best
linear fit for the FRIs+FRIIs. This is also true for the
subsamples in Fig.~\ref{fig4}, except the PG quasars which is
close to slope=1 but with larger errors in Table~\ref{tab1}. The
correlation coefficient and confidence level are quite high for
the radio galaxies, Seyferts+LINERS in the two samples in
Table~\ref{tab2}. We did not fit for the radio loud quasars
(RLQs), because they are clustering in a small area making it
difficult to fit them properly.

\begin{table}
         \caption[]{The linear fit parameters of $log[P_{j}]=b+\mu \times log[\lambda m]$ along with the derived value of
$\eta^{'}/\varepsilon^{'}$ in the upper part, and the linear fit
parameters of $log[R]=c+\rho \times log[\lambda m]$ in the lower
part, for narrow line radio galaxies (NLRG), radio galaxies (RG),
Seyfert galaxies and LINERS (Sey+L), and PG quasars (PGQ),
respectively. The number in the brackets of second column is the
coefficient of determination (COD) which ranges from 0-1, it
basically equals to the square of the Pearson coefficient of
linear correlation. The error is 2$\sigma$ (standard deviation
from the linear fit program).}
         \label{tab1}
\small

         $$
         \begin{tabular}{c|c|c|c|c }

            \hline
            \noalign{\smallskip}

Sample & subsample   & $\mu$ & $b$ & $\eta^{'}/\varepsilon^{'}$\\

            \noalign{\smallskip}
            \hline
            \noalign{\smallskip}

Sikora13 & NLRG(0.52) & 0.52$\pm$0.06 & 40.10$\pm$0.34 & 100.0  \\

            \hline

Sikora07 & RG(0.51) & 0.35$\pm$0.10 & 41.06$\pm$0.60 & 912.0 \\

& Sey+L(0.47) & 0.32$\pm$0.12 & 38.32$\pm$0.58 & 1.7 \\

& PGQ(0.47)  & 0.92$\pm$0.44 & 34.08$\pm$3.42 & 9.5E-5 \\
   \hline

 &   & $\rho$ & $c$ & \\
  \hline

Sikora13& NLRG(0.32)  & -0.40$\pm$0.08 & 3.46$\pm$0.40 &   \\
   \hline

Sikora07& RG(0.69)  & -0.59$\pm$0.12 & 7.22$\pm$0.70 &   \\

& Sey+L(0.73)  & -0.63$\pm$0.14 & 4.02$\pm$0.66 &   \\

& PGQ(0.00)  & 0.07$\pm$0.52 & -0.89$\pm$3.96 &   \\

             \noalign{\smallskip}
            \hline

           \end{tabular}{}
         $$
   \end{table}

\begin{table}
         \caption[]{The Pearson, Spearman and Kendall linear correlation coefficient between $log[P_{j}]$ and $log[\lambda m]$, and
         that between $log[radio loudness]$ and $log[\lambda m]$ marked `logR' in second line, with
         null hypothesis probability (non correlation probability) in
         brackets.}

         \label{tab2}
\small
         $$
         \begin{tabular}{c|c|c|c}

            \hline
            \noalign{\smallskip}

 Subsample  &  Pearson & Spearman & Kendall \\

            \noalign{\smallskip}
            \hline
            \noalign{\smallskip}
 NLRG  & 0.72(0)  & 0.71(0)  &  0.53(0)  \\

 logR &  -0.57(0) & -0.52(0)  &  -0.36(1.6E-18)   \\

            \hline

 RG &  0.71(5.7E-9) &  0.64(6.4E-7)  & 0.45(5.7E-6)   \\
 logR &  -0.83(1.1E-13) & -0.81(1.1E-12)  &  -0.62(2.5E-10)   \\

 Sey+L &0.68(4.2E-6) &  0.68(5.2E-6) & 0.51(1.5E-5)   \\
 logR &  -0.85(3.6E-11) & -0.86(2.4E-11)  &  -0.70(2.2E-9)   \\

 PGQ &  0.69(5.5E-4)  & 0.64(0.002) & 0.57(3.9E-4)  \\
 logR &  0.06(0.78) & 0.17(0.45)  &  0.17(0.29)   \\

             \noalign{\smallskip}
            \hline

           \end{tabular}{}
         $$
   \end{table}

In the results, we show that the jet powers at 1.4 GHz positively
correlate with the $\lambda m$, but not linearly, i.e. the power
law indices ($\mu$) are significantly less than unity for the
radio galaxies and Seyferts+LINERS. The radio loudness is
anti-correlated with the $\lambda m$ for the radio galaxies and
Seyferts+LINERS. For PG quasars, there appears to be a linear
proportionality between the jet power and disc luminosity but with
larger errors, that leads to no correlation between radio loudness
and disc luminosity.

To further study the anti-correlation between the radio loudness
and the $\lambda m$ we found, considering $P_{j}=\eta^{'}
/\varepsilon^{'} L_{disc}^{\mu}$ and equation (8)-(9), we have
radio loudness:

\begin{equation}
R=P_{1.4}/L_{line} \propto P_{j}^{7/6}/L_{disc} \propto
L_{disc}^{(7/6)\mu-1}\propto L_{disc}^{\rho}
\end{equation}

The ratio $\eta^{'} /\varepsilon^{'}$ can be estimated from the
$b$ of the linear fit parameters in Table~\ref{tab1}. From
equation (12) we find a relation $\rho=(7/6)\mu-1$. This relation
accords with the fitting results in Table~\ref{tab1}, e.g. for
$\mu=0.52$, the $\rho=(7/6)\mu-1$= -0.39 is close to the measured
value $\rho=-0.40\pm0.08$ in Table~\ref{tab1} for the NLRGs. So
the anti-correlation is apparently explained, that is due to the
$\mu<6/7$, so that $\rho=(7/6)\mu-1<0$.

The physics for $\mu<6/7$ (0.86) needs to be explored further in
the $P_{j}= (\eta^{'} /\varepsilon^{'}) L_{d}^{<0.86}$, it implies
that the jet power increases less efficiently than the disc
luminosity increases. For $\mu<1$, the $\eta$ and $\varepsilon$
may be not constants but vary with the accretion rate.
\citet{sik13} used the relation of
$\varepsilon\propto\lambda^{2/5}$ for the BH magnetosphere of a
truncated disc, in which the jet power is dominated by a BH spin.
With $R=P_{1.4}/L_{line}\propto\eta /\varepsilon$ approximately,
they explain that the anti-correlation in radio loudness and
Eddington ratio could be due to the
$\varepsilon\propto\lambda^{2/5}$, i.e. $R\propto\lambda^{-0.4}$.

\citet{van13} find that from 763 FRII quasars with the median
redshift of 1.16, a linear correlation between 1.4 GHz luminosity
and bolometric disc luminosity with $P_{1.4}\propto L_{bol}$. If
we use the equation (8), the jet power will be $P_{j}\propto
P_{1.4}^{6/7} \propto L_{bol}^{0.86}$. They claim that, for this
nearly linear proportionality, the power output from the inner
part of the accretion disc dominates over the power extracted from
the black hole by the Blandford-Znajek mechanism (BZ-jet,
hereafter).

\citet{kal12} investigated the $[O_{II}]$ emission line properties
of 18508 quasars at z $<$ 1.6 drawn from the Sloan Digital Sky
Survey (SDSS) quasar sample. The quasar sample has 1692 radio-loud
(RLQs) and 16816 radio-quiet quasars (RQQs), according to the
traditional radio-loud/quiet division of the radio-to-optical flux
ratio of 10, and the radio loudness computed using the radio flux
density at 1.4 GHz and the optical (7480$\AA$) flux density from
the SDSS. They found a strong correlation between 1.4 GHz radio
luminosity and narrow emission-line luminosity, with a power law
index of $\mu\sim1.6$ and $\sim$1.2 for the RLQs and RQQs
respectively. For radio jet power, using the equation (8), the
slope reduces to $\mu\sim1.4$ and $\sim$1.0 for the RLQs and RQQs
respectively. The optical line luminosity is linearly proportional
to the disc luminosity in \citet{kal12}.

\section{Discussion}

From the results above, in general there is a positive power law
correlation between radio jet power and disc luminosity
$P_{j}=\eta^{'} /\varepsilon^{'} L_{disc}^{\mu}$. The power law
indices are significantly less than unity for the Seyfert
galaxies, FRI and FRII galaxies at z $<$ 0.5 and the Eddington
ratio $\lambda<0.1$ (or typically $\lambda m<7$ in Fig.~\ref{fig1}
and Fig.~\ref{fig4}). Second, there is an anti-correlation between
radio loudness and disc luminosity for the samples which have the
$\mu<$6/7. We find that the negative correlation is caused by
$(7/6)\mu <1$ with the relation $\rho=(7/6)\mu-1<0$. On the other
hand, in high-z quasars from the literature, we have $\mu\sim0.86$
for the FRII quasars in \citet{van13} and $\mu>1$ for the quasars
in \citet{kal12}. It will lead to a positive correlation between
radio loudness and disc luminosity for $\mu>6/7$ and no
correlation for $\mu\sim6/7$ in the high-z samples, and we will
investigate these in future.

The radio loudness problem appears to be resolved with our
interpretation, the underlying physics is still not clearly
identified. Are there different accretion discs or jet forming
mechanisms for the power-law correlation index $\mu\ll$1 and
$\mu\geq$1? In the BH accretion models, there are cold and hot
accretion flows. Cold accretion flow consists of optically thick
and geometrically thin gas, e.g. the standard thin disc
\citep{shak73}, which occurs at a fraction of the Eddington mass
accretion rate, and the slim disc at super-Eddington rates
\citep{abr88}. Hot accretion flows, however, are virially hot and
optically thin, they occur at lower mass accretion rates, and are
described by models such as the advection-dominated accretion flow
(ADAF, \citealt{nar94}) and luminous hot accretion flow.
Observations show that hot accretion flows are often associated
with jets, and they are present in low-luminosity AGN (LLAGN) (see
\citealt{yuan14}, for a review on hot accretion flows).
\citet{ho01} and \citet{ho02} find that most of the LLAGN in
nearby galaxies are radio-loud, and the radio loudness is
inversely correlated with the disc luminosity. \citet{sik13} and
\citet{sikb13} find that the anti-correlation of radio loudness
and disc luminosity could be explained with the
$\varepsilon\propto\lambda^{2/5}$ in the BH spin paradigm,
implying that the BH spin plays an important role in the jet
powers. Therefore the Seyfert galaxies and radio galaxies in our
samples are likely powered by both the inner part of accretion
disc and the BH spin. If the BZ-jet power is gradually suppressed
when the accretion rate increases, the jet power could be less
proportional to the disc luminosity and this may lead to an
anti-correlation of radio loudness and disc luminosity. This is
some similar to the transition from hard/low (with low accretion
rate and radio jet) to soft/high (with high accretion rate and no
jet) state in the X-ray binaries, in which a contribution of jet
power from black hole spin is possible \citep{fen04}. However,
this explanation needs a high fraction of rapid spin BHs in the
low-z AGN.

\citet{mar11a, mar11b} analyzed the ratio of jet power to disc
luminosity, they find that at z$<$0.5 the low-excitation galaxies
have low accretion rates and bimodal spin distributions, with
approximately half of the population having maximal spins, while
high exciting galaxies are explained as high-accretion rate but
very low spin objects at higher redshifts (z$\sim$1) and only a
small population of nearly maximally spinning high accretion rate
objects is possible. This may be supported by the findings that
much higher fraction of radio loud galaxies at low accretion rates
\citep{ho01, ho02} than the fraction ($\sim$10\%) of radio loud
quasars at higher redshifts \citep{kal12}. And it has been
suggested that low redshift FRIs have rapid spinning BHs
\citep{wu11}.

In high-z quasars, a steep slope $\mu \sim 1$ or $>$1 between jet
power and disc luminosity is often observed, e.g, \citet{van13},
\citet{wil99}, \citet{fer11}, \citet{kal12}. The quasars are at
high accretion state ($\lambda>0.1$) and hosted by elliptical
galaxies. The massive ellipticals are most likely formed via major
merger events, and their nuclei may have different nuclear
environments than disc galaxies. \citet{fal10} find evidence for
the environmental source density to increase with the radio
luminosity of AGN at around z=1. The jets of FRII quasars are
probably launched through the Bondi accretion of hot interstellar
gas \citep{bon52, all06, gas13, sik13}, with efficient jet powers
\citep{wer12}. \citet{all06} find a tight positive correlation
between the Bondi accretion rate and the radio power required to
inflate cavities observed in the surrounding X-ray emitting gas,
suggests that the Bondi formulae provide a reasonable description
of the accretion process for powerful jets. Furthermore, the
powerful jet may live in a short time for quasars, given that the
majority of quasars are radio quiet.

\citet{mei01} proposed a hybrid model combined both the disc
accretion and the BH spin/magnetosphere effects, it can produce
powerful jets if the BH spin $a_{\ast}>0.9$ ($a_{\ast}$ is
dimensionless spin parameter ranging from 0-1, see also
\citealt{nem07}). Therefore, both the hybrid model and the
magnetically arrested disc \citep{tch11, mck12} needs maximally
spinning BHs for producing FRII jets, it is suggested that only a
small population of nearly maximally spinning BHs in high
accretion rate objects, and spins of BHs are generally low at
around z=1 \citep{mar11a, mar11b}. As noted by \citet{van13}, the
powerful jets of FRII quasars are mainly controlled by the inner
part of accretion flows rather than by the power extracted from
the BH spin (also see \citealt{liv99}), in contrast to that the
BZ-jet may control the negative correlation of radio loudness and
disc luminosity in the low luminosity AGN as we discussed above.

\citet{fer11} find from a radio selected sample of 27 radio
galaxies at a narrow redshift range z=0.9-1.1 that is unbiased to
evolutionary effects, that there is a tight positive correlation
between the radio luminosity and the mid-IR (also the $[O_{II}]$)
line luminosity with power law indices $\mu>1$. They added optical
selected quasars (OSQs) into the analysis, the correlation for the
brightest radio sources appears to become an upper envelope.
However, \citet{kal12} investigated both radio loud and quiet
quasars at z $<$ 1.6 drawn from the SDSS quasar sample. They find
a strong correlation between 1.4 GHz radio luminosity and narrow
emission-line luminosity, for both RLQs and RQQs, with power law
indices $\mu>1$. This sample is much larger than the \citet{fer11}
sample, so it could be statistically more significant. Our studies
concentrate mainly on radio loud sources, and it is needed to
include radio quiet ones in larger and complete samples into
analysis in future.

Finally, we note that the different slopes of correlation between
radio jet power and disc luminosity might be affected by selection
effects, e.g. sample sizes etc., however, we have high confidence
for that in high-z quasars the slopes are much steeper than those
of low redshift galaxies in our analysis. There are no correlation
found between radio luminosities and source sizes in \citet{wil99}
and \citet{sik13} for FRIs and FRIIs, however, caution must be
taken if there is a wide range of source sizes in a relative small
sample \citep{sha13}. The Doppler boosting effect is not
considered here, since our samples are mainly the FRIs, FRIIs and
Seyfert galaxies, and the blazars are not included. Furthermore,
as noted by \citet{sing14}, the radio power and/or the disc
luminosity may be also related to the redshift, a redshift --
luminosity correlation (Malmquist bias) in a flux-limited sample.
We checked our data, find there are weak positive correlations for
the radio galaxies while stronger correlation for the PG quasars
between the radio jet power (or disc luminosity) and redshift. In
fact, the correlation is stronger between the radio power and the
disc luminosity than that between the radio (or disc) luminosity
and redshift in the radio galaxies of our samples, these type of
source are mainly studied in this paper, but the Malmquist bias
might still have some effects which we were currently not able to
remove from our data. As analyzed by \citet{sing14}, a flux
limited sample with a wide range of radio luminosities within
narrow redshift ranges (and vice versa) will be needed to
disentangle the effects in future.

\section{Summary}

A model for the relation between radio jet power and the product
of central BH mass and Eddington ratio of AGN is proposed. The
model is examined with data from the literature.

We find that radio jet power positively correlates but not
linearly with the product of BH mass and Eddington ratio, and the
power law indices are significantly less than unity for relatively
low accretion ($\lambda<0.1$) AGN, $P_{j}\propto (\lambda
m)^{\mu}$, in the radio galaxies and the Seyfert galaxies. This
leads to a negative correlation between radio loudness and
$\lambda m$ for the low luminosity AGN, i.e. $R\propto (\lambda
m)^{\rho}$ with $\rho=(7/6)\mu-1<0$, which may be attributed to a
contribution of BZ-jet to total jet power assuming that the BZ-jet
power is gradually suppressed as the accretion rate increases.

On the contrary, for the high-z quasars which often show the slope
$\mu\geq1$, a positive correlation between the radio loudness and
disc luminosity is predicted. We discuss that the jet powers of
the high-z FRII quasars are likely dominated by the accretion disc
rather than by the BH spin.

\section*{Acknowledgments}

We thank the reviewer for useful comments. This work is supported
by the National Natural Science Foundation of China (Grant
No.11273050) and the 973 Program of China (2009CB824800). This
research has made use of the NASA/IPAC Extragalactic Database
(NED) which is operated by the Jet Propulsion Laboratory,
California Institute of Technology, under contract with the
National Aeronautics and Space Administration.

\end{document}